\documentclass[a4paper,fleqn,10pt]{article}
\usepackage[numbers,sort&compress]{natbib}
\usepackage{graphicx}
\usepackage{latexsym}
\usepackage{amsmath,amssymb}
\usepackage{dcolumn}
\usepackage{float}
\usepackage{caption}

\begin{document}
\captionsetup[figure]{labelfont={bf},labelformat={default},labelsep=space,name={Fig.}}

\title{{\bf Analytical time-like geodesics in modified Hayward black hole space-time}
\author{\normalsize Jian-Ping Hu$^{1}$, Li-Li Shi$^{1}$,
       Yu Zhang$^{1}$\thanks{Corresponding Author(Y. Zhang): Email: zhangyu\_128@126.com},
        and Peng-Fei Duan$^{2}$ \\
   \normalsize \emph{$^{1}${Faculty of Science, Kunming University of Science and Technology, }}\\ \normalsize \emph {Kunming, Yunnan 650500, People's Republic of China}\\
   \normalsize \emph{$^{2}$Geological Resources and Geological Engineering Postdoctoral Programme, Kunming University of} \\ \normalsize \emph {Science and Technology, Kunming, Yunnan 650093, People's Republic of China}}}

\date{}
\maketitle \baselineskip 0.3in

\begin{abstract}
The properties of modified Hayward black hole space-time can be investigated through analyzing the particle geodesics. By means of a detailed analysis of the corresponding effective potentials for a massive particle, we find all possible orbits which are allowed by the energy levels. The trajectories of orbits are plotted by solving the equation of orbital motion numerically. We conclude that whether there is an escape orbit is associated with $b$ (angular momentum). The properties of orbital motion are related to $b$, $\alpha$ ($\alpha$ is associated with the time delay) and $\beta$ ($\beta$ is related to 1-loop quantum corrections). There are no escape orbits when $b$ $<$ $4.016M$, $\alpha$ = 0.50 and $\beta$ = 1.00. For fixed $\alpha$ = 0.50 and $\beta$ = 1.00, if $b$ $<$ $3.493M$, there only exist unstable orbits. Comparing with the regular Hayward black hole, we go for a reasonable speculation by mean of the existing calculating results that the introduction of the modified term makes the radius of the innermost circular orbit (ISCO) and the corresponding angular momentum larger.
\end{abstract}


\section{Introduction}
Einstein's general relativity is broken down at singularities.
A black hole without singularity is called a regular black hole or a nonsingular black hole; the first regular model was presented by Bardeen\cite{bardeen1968non}.
Since then, many other regular models \citep{garfinkle1991charged,ayon-beato1998regular,wu2008general,horava2009quantum}are presented. For example, the regular Hayward black hole was presented by Hayward\cite{hayward2006angular}, a new parameter $\ell$ (a convenient encoding of the central energy density $3/8\pi\ell^{2}$) was considered in this model.
The metric of the regular Hayward black hole is as follows \citep{hayward2006angular}
\begin{eqnarray}
&ds^{2}=&-(1-\frac{2Mr^{2}}{r^{3}+2Ml^{2}})dt^{2}+\frac{1}{1-\frac{2Mr^{2}}{r^{3}+2Ml^{2}}}dr^{2}+r^{2}d\theta^{2}+r^{2}\sin^{2}\theta d\phi^{2}
\label{e1}
\end{eqnarray}
But the regular model is not perfect.
Some modifications could be introduced to improve the regular metric.
De Lorenzo et al.\cite{de lorenzo2015onthe} proved taking into account a time delay and the 1-loop quantum correction can improve the regular Hayward metric in the central core.
The modified Hayward metric is as follows\citep{de lorenzo2015onthe}
\begin{eqnarray}
&ds^{2}=&-(1-\frac{2Mr^{2}}{r^{3}+2Ml^{2}})(1-\frac{\alpha \beta M}{\alpha r^{3}+ \beta M})dt^{2}+\frac{1}{1-\frac{2Mr^{2}}{r^{3}+2Ml^{2}}}dr^{2}+r^{2}d\theta^{2}+r^{2}\sin^{2}\theta d\phi^{2}
\label{e2}
\end{eqnarray}
 Here, ($1-\frac{\alpha \beta M}{\alpha r^{3}+ \beta M}$) is the correcting factor. The constant $\alpha$ is associated with the time delay between the center and infinity. From $(\delta t_{\infty}-\delta t_{0})$/$\delta t_{\infty}=1-\sqrt{|g_{00}(r=0)|}\in[0,1)$, it is obtained that the values of $\alpha \in [0,1)$, and the larger $\alpha$, the greater the time delay. The parameter $\beta$ is related to the 1-loop quantum corrections of the Newtonian potential, the suggested maximum value is given in Ref. \citep{de lorenzo2015onthe}, $\beta_{max}$=41/(10$\pi$). When $\alpha$ =0 or $\beta$=0, the metric of modified Hayward will revert to the regular Hayward metric.
 The detailed work is shown in Ref. \citep{de lorenzo2015onthe}. He discussed bounds of the maximal time delay caused by curvature conditions, and the consequences for the weak energy condition. Up to now, there has been much research\citep{debnath2015accretion,pourhassan2016effects,zhao2017strong,perez-roman2018theregion} on modified Hayward black hole space-time. Debnath\cite{debnath2015accretion} has analyzed the accretion of the fluid flow on the modified Hayward black hole , then he calculated the critical point, the fluid's four-velocity, and the velocity of sound during the accretion process.
 He also analyzed the evaporation of this black hole.
 Pourhassan et al.\citep{pourhassan2016effects} have investigated the impacts of thermal fluctuations on modified Hayward black hole thermodynamics, and found that the modified Hayward black hole is stable even after the thermal fluctuations are taken into account, if only the event horizon is larger than a certain critical value.
 The study of observables (angular separations, brightness differences and time delays between its relativistic images) in the strong deflection gravitational lensing\citep{virbhadra2000schwarzschild,virbhadra2008relativistic,chen2009strong,sadeghi2013strong,ji2014strong,younas2015strong} by a modified Hayward black hole has been discussed in Zhao and Xie\citep{zhao2017strong}.
 They proposed that it is likely to distinguish the modified Hayward black hole from a Schwarzschild one, but it needs a very high resolution.
 Their work made a better comprehension on the properties of the modified Hayward black hole for us.

 In this paper, we will explore the trajectories of a massive particle in modified Hayward black hole space-time. As we all known, the black hole is a unique object which has super strong gravity near the center core.
 The information inside the horizon of a black hole cannot be detected. However we can still understand the geometric structure of a black hole by studying the trajectories of a massive particle or photon outside the event horizon, and the study of geodesic structure\citep{stuchlik1991null,cruz1994geodesic,beem1997stability,chandrasekhar1998mathematical,breton2002geodesic,teo2003spherical,cruz2005geodesic,cardoso2009geodesic,chiba2017anote,dasgupta2009kinematics,abdujabbarov2010test,kagramanova2010analytic,grunau2011geodesics,grunau2012geodesic,kagramanova2012geodesic,kostic2012analytical,grunau2013geodesic,abbas2014geodesic,chakraborty2014inner,zhang2014orbital,soroushfar2015analytical,uniyal2015geodesic,uniyal2015geodesics,kuniyal2016null,shenavar2016motion,soroushfar2016geodesic,al-badawi2017geodesics,azam2017geodesic,azam2017geodesics,sharif2017motion,ghaderi2017geodesics,uniyal2017nullgeodesics} is one of the hot topics in black hole physics. By analyzing the corresponding effective potential for particle and photon, Cruz et al.\cite{cruz2005geodesic} found all types of orbital motion which are permitted by the energy levels in Schwarzschild Anti-de Sitter black hole space-time.
 Their results showed that if the constants of motion satisfies the condition $E^{2}<L^{2}/\ell^{2}$, there exist bounded orbits which not exist in the Schwarzschild space-time. In Ref. \citep{kostic2012analytical}, the time-like orbits were divided into four types (scattering orbits, plunging orbits, near orbits and bound orbits) in the Schwarzschild space-time by mean of Jacobi elliptic functions and elliptic.
 The radial and circular trajectories were investigated in Ref. \citep{al-badawi2017geodesics} by analyzing the behavior of effective potentials for the massive and massless particle. It was obtained that an exact analytical solution for ISCO and the radius of ISCO shrinks due to the presence of electromagnetic field.
 Recently, the trajectories of photon motion in the background of Kerr-Sen black hole arising in heterotic string theory were investigated by Uniyal et al.\cite{uniyal2017nullgeodesics}.
 They also calculated the rotation and mass parameters for the Kerr-Sen black hole and analyzed observables on the angular plane.

In this short note, we investigate the time-like geodesic in the modified Hayward black hole space-time by solving the equation of orbital motion. We prove the influence of modifications on the effective potential, and contrast the geodesic structures of the regular Hayward black hole space-time with the modified Hayward black hole space-time ones.
The structure of this paper is organized as follows: In Sect.2, we obtain the equation of orbital motion and give the effective potential. In Sect.3, we discuss the time-like geodesic structure of modified Hayward black hole space-time in detail. In the concluding section, a brief conclusion is given.


\section{Equation of orbital motion and effective potential}
\begin{figure}[ht!]
\centering
    \includegraphics[angle=0, width=0.4\textwidth]{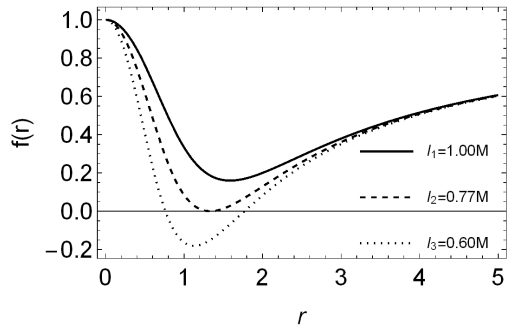}
     \vspace*{8pt}
\caption{Horizons of the modified Hayward black hole space-time with different values of $\ell$ \label{v1} }
\end{figure}

In Eq. (\ref{e2}), $1-\frac{2Mr^{2}}{r^{3}+2Ml^{2}}$ is the lapse function of the modified Hayward black hole space-time which is the same as that of the regular Hayward black hole space-time, so they have the same horizon types.
By analyzing the lapse function, we can distinguish three kinds of different space-time: no horizon (${\frac{\ell^{2}}{M^{2}}}>\frac{16}{27}$), one horizon (${\frac{\ell^{2}}{M^{2}}} = \frac{16}{27}$) and double horizons (${\frac{\ell^{2}}{M^{2}}} < \frac{16}{27}$) which is shown in Fig. \ref{v1}.

The variation of the Euler-Lagrange equation associated with the metric can be adopted to describe the geodesic structure. The form of the corresponding Lagrangian for a massive particle according to the metric of the modified Hayward black hole is written as
\begin{eqnarray}
 &L=&\frac{1}{2}m(\frac{ds}{d\tau})^2=-\frac{1}{2}mf(r)B(r)\dot{t}^2+\frac{1}{2}m\frac{1}{B(r)}\dot{r}^2+\frac{1}{2}mr^2\dot{\theta}^2+\frac{1}{2}mr^{2}\sin^{2}\theta\dot{\phi}^2,\nonumber
\end{eqnarray}
\begin{eqnarray}
&B(r)=1-\frac{2Mr^{2}}{r^{3}+2Ml^{2}}, \ \ \ f(r)=1-\frac{\alpha \beta M}{\alpha r^{3}+ \beta M}.
\label{e3}
\end{eqnarray}
In here, $m$ is the mass of the test particle and $\tau$ is the proper time. $\dot{t}$ = $dt/d\tau$,\  $\dot{r}$ = $dr/d\tau$, \ $\dot{\theta}$ = $d\theta/d\tau$, \ $\dot{\phi}$ = $d\phi/d\tau$. Without losing generality, we choose $\theta$ = $\frac{\pi}{2}$, $\dot{\theta}$ = 0, a new expression for Eq. (\ref{e3}) is given by
\begin{eqnarray}
&L=-\frac{1}{2}mf(r)B(r)\dot{t}^2+\frac{1}{2}m\frac{1}{B(r)}\dot{r}^2+\frac{1}{2}mr^{2}\dot{\phi}^2.
 \label{e4}
\end{eqnarray}
\begin{eqnarray}
&\frac{d}{d\tau}(\frac{\partial L}{\partial \dot{x}^{\nu}})-\frac{\partial L}{\partial x^{\nu}}=0.
\label{e5}
\end{eqnarray}
Considering Eq. (\ref{e4}) does not explicitly contain $t$ and $\phi$, by using Eq. (\ref{e5}), we get two equations as follows
\begin{eqnarray}
&\frac{\partial L}{\partial t}=0 \Longrightarrow{-\frac{\partial L}{\partial \dot{t}}=\varepsilon=m f(r) B(r)\dot{t}},
\label{e6}
\end{eqnarray}
\begin{eqnarray}
&\frac{\partial L}{\partial \phi}=0 \Longrightarrow{-\frac{\partial L}{\partial \dot{\phi}}=J=m r^{2}\dot{\phi}}.
\label{e7}
\end{eqnarray}
By analyzing Eqs. (\ref{e6}) and (\ref{e7}), we obtain two constants, the total energy $\varepsilon$ and the total angular momentum $J$.\\
Define $E$ = $\frac{\varepsilon}{m}$, \ $b$ = $\frac{J}{m}$ and bring it to Eqs. (\ref{e6}) and (\ref{e7}), we get
\begin{eqnarray}
&\dot{t}=\frac{E}{f(r)B(r)}, \ \ \ \dot{\phi}=\frac{b}{r^{2}},
\label{e8}
\end{eqnarray}
then we can obtain the equation of orbital motion
\begin{eqnarray}
&\dot{r}^{2}=\frac{E^{2}}{f(r)}-(h+\frac{b^{2}}{r^{2}})B(r).
\label{e9}
\end{eqnarray}
According to Eq. (\ref{e9}), the effective potential can be defined as
\begin{eqnarray}
& V_{eff}^{2}=f(r)B(r)(h+\frac{b^{2}}{r^{2}}).
\label{e10}
\end{eqnarray}

\section{Time-like geodesic structure}
\subsection{Effect of modified term on the stability of orbital motion}
For massive particle, $h$ = 1, the corresponding effective potential equation is written as
\begin{eqnarray}
 &V_{eff}^{2}=(1-\frac{2Mr^{2}}{r^{3}+2Ml^{2}})(1-\frac{\alpha \beta M}{\alpha r^{3}+ \beta M})(1+\frac{b^{2}}{r^{2}}).
\label{e11}
\end{eqnarray}

 \begin{figure}[ht!]
\centering
    \includegraphics[angle=0, width=0.3\textwidth]{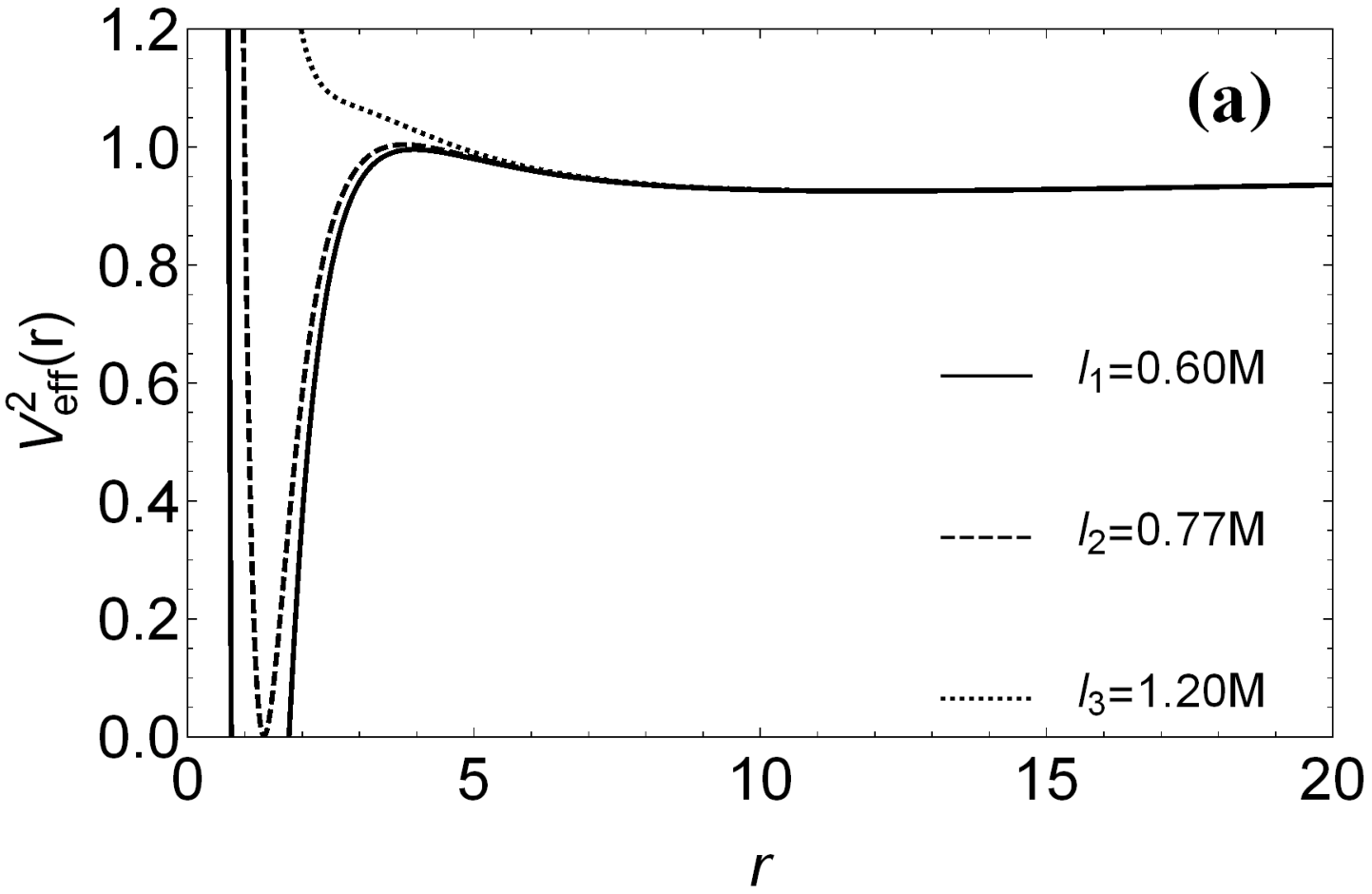}
    \includegraphics[angle=0, width=0.3\textwidth]{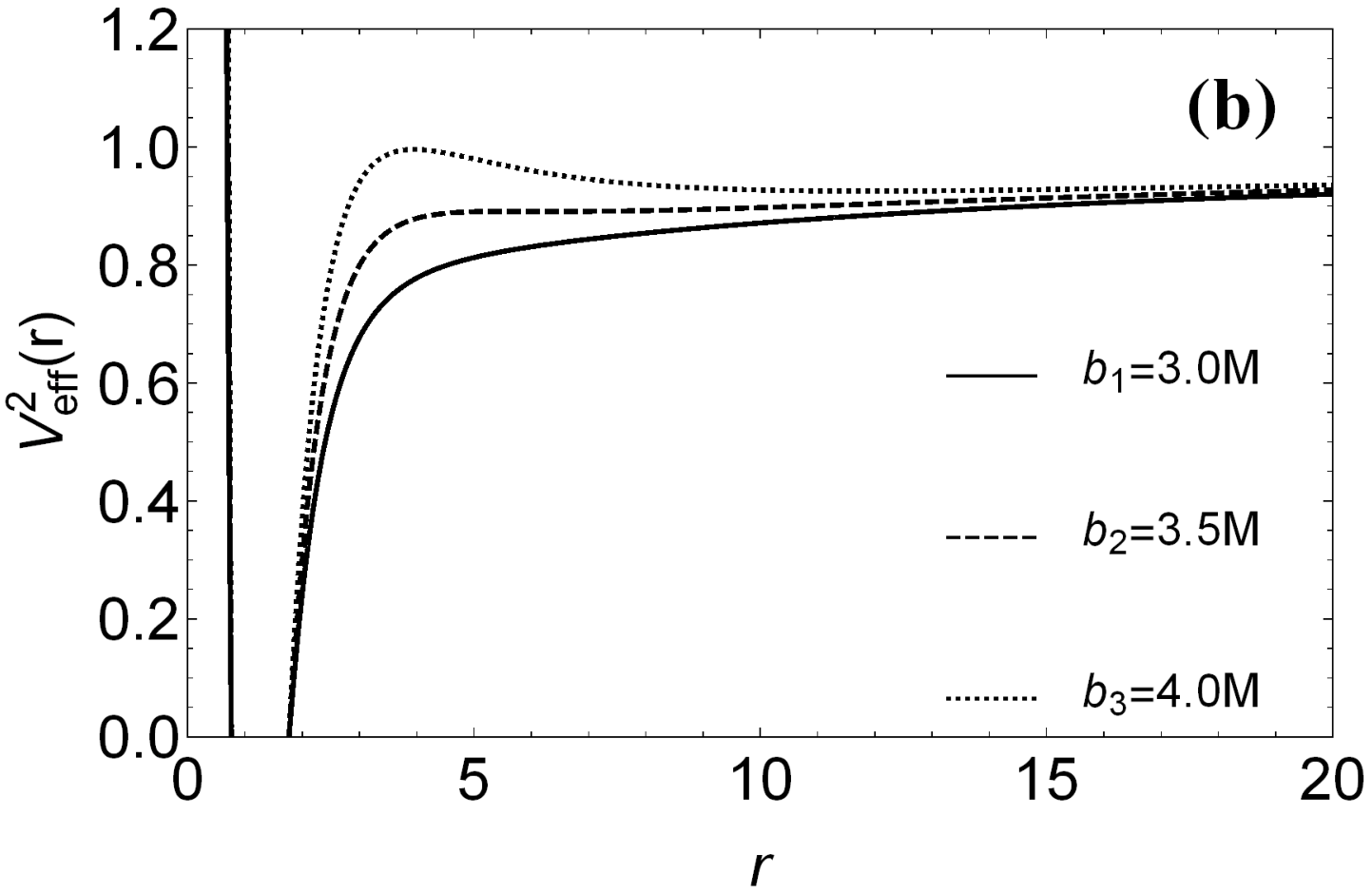} \\
     \includegraphics[angle=0, width=0.3\textwidth]{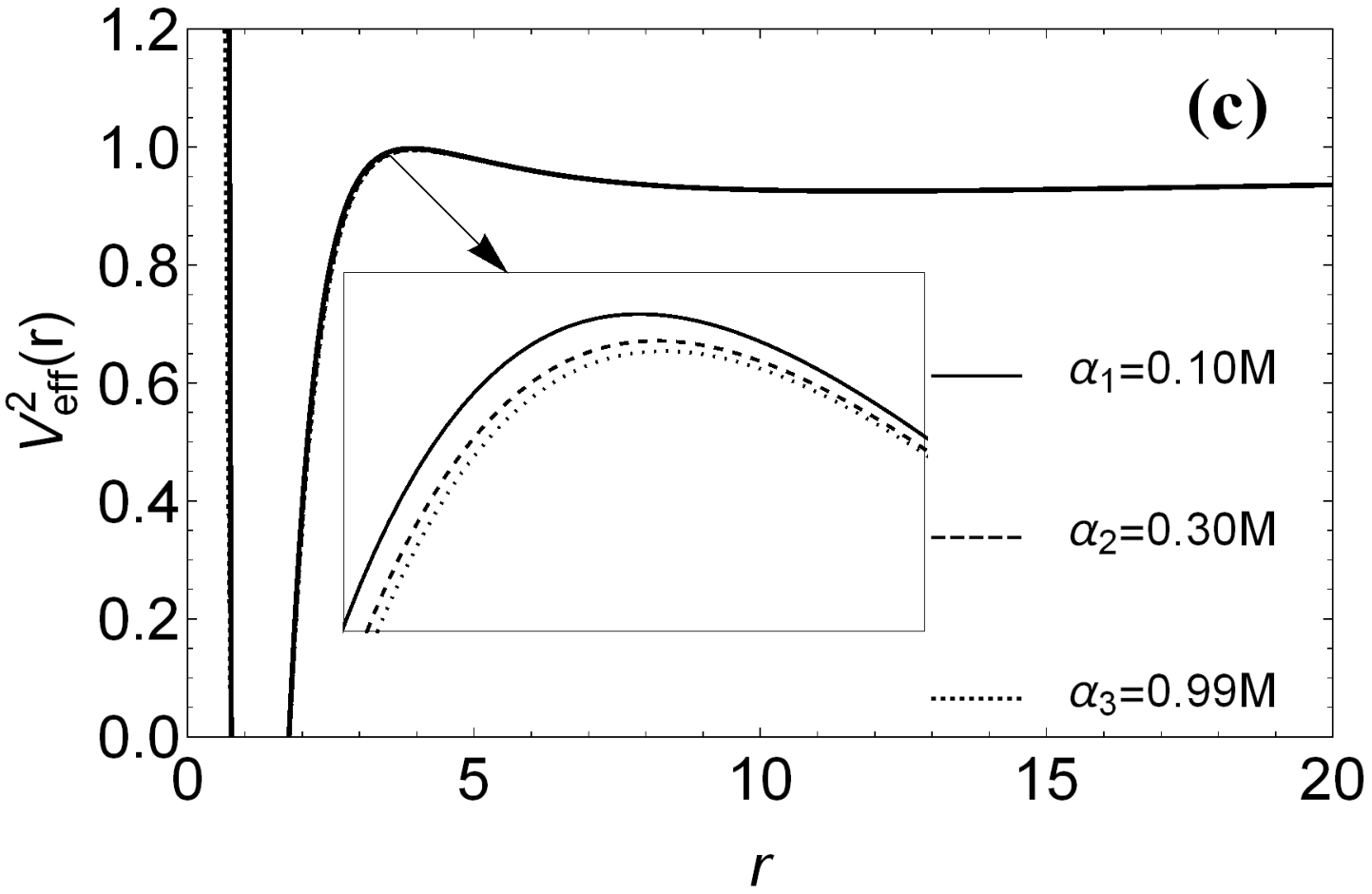}
    \includegraphics[angle=0, width=0.3\textwidth]{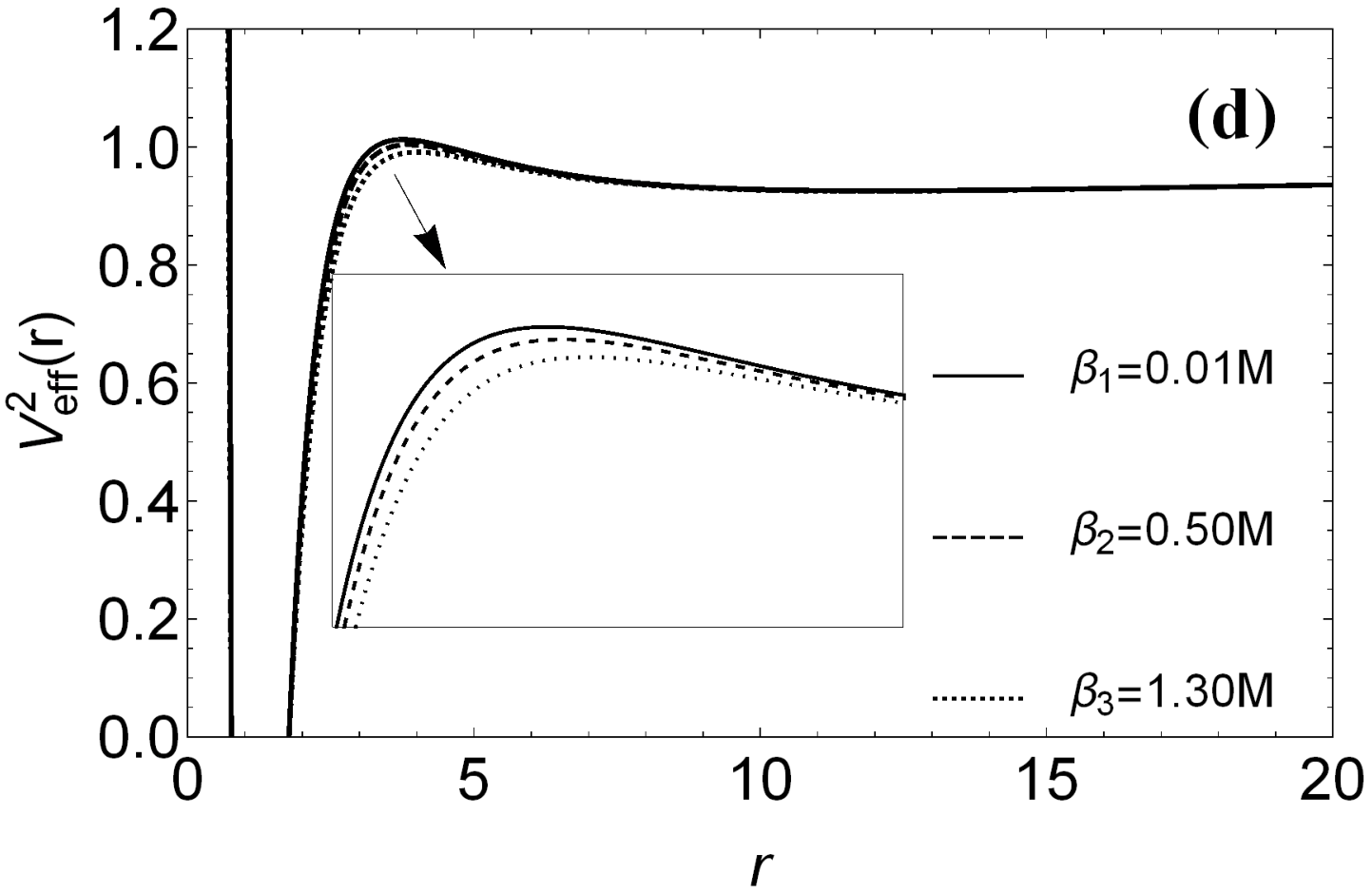}
     \vspace*{8pt}
\caption{The behaviors of the effective potential for the massive particle with different values of parameters $\ell$, $b$, $\alpha$ and $\beta$ which correspond to (a), (b), (c) and (d), respectively.\label{v2} }
\end{figure}
\begin{figure}[ht!]
\centering
    \includegraphics[angle=0, width=0.35\textwidth]{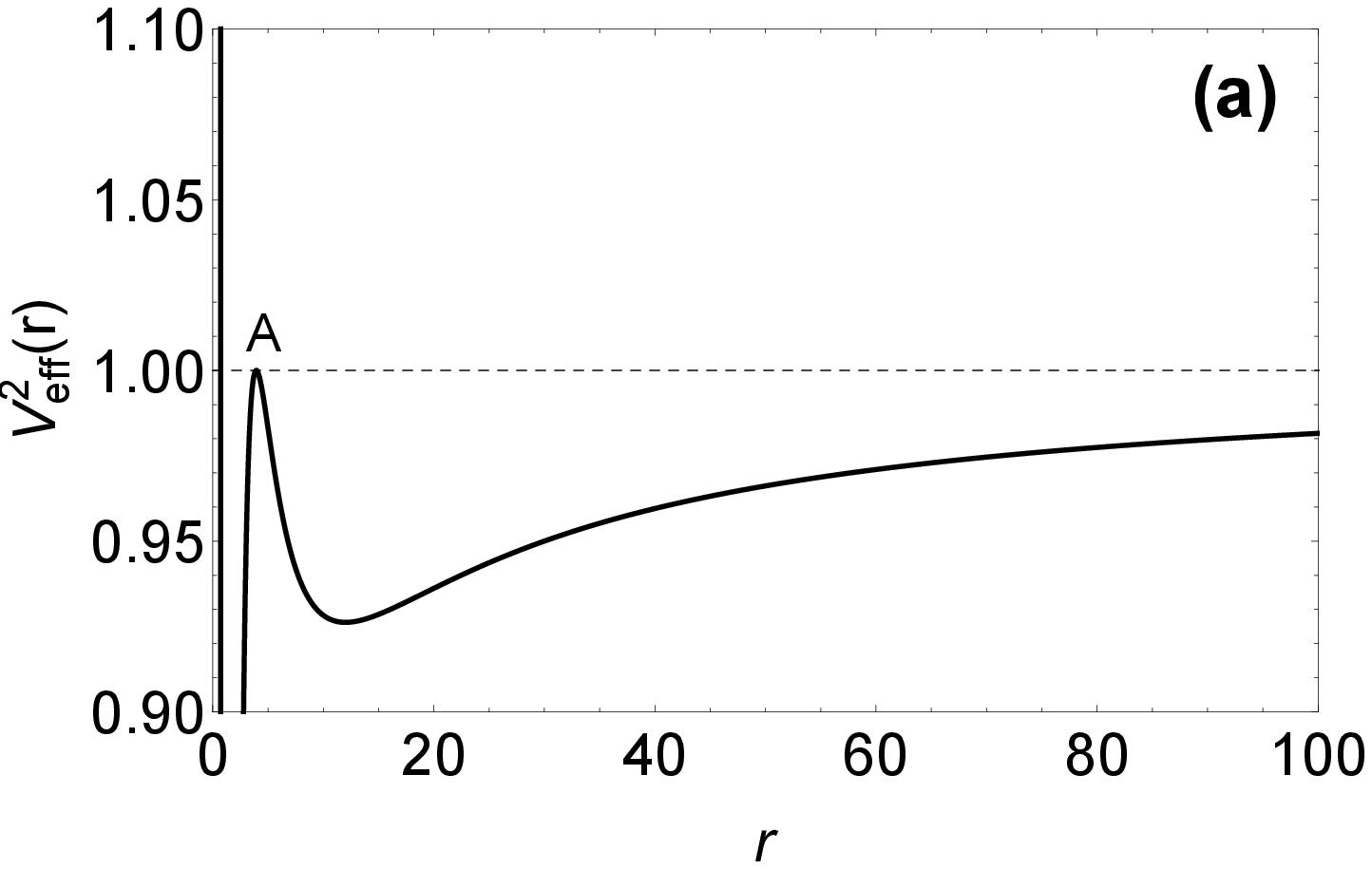}
    \includegraphics[angle=0, width=0.3\textwidth]{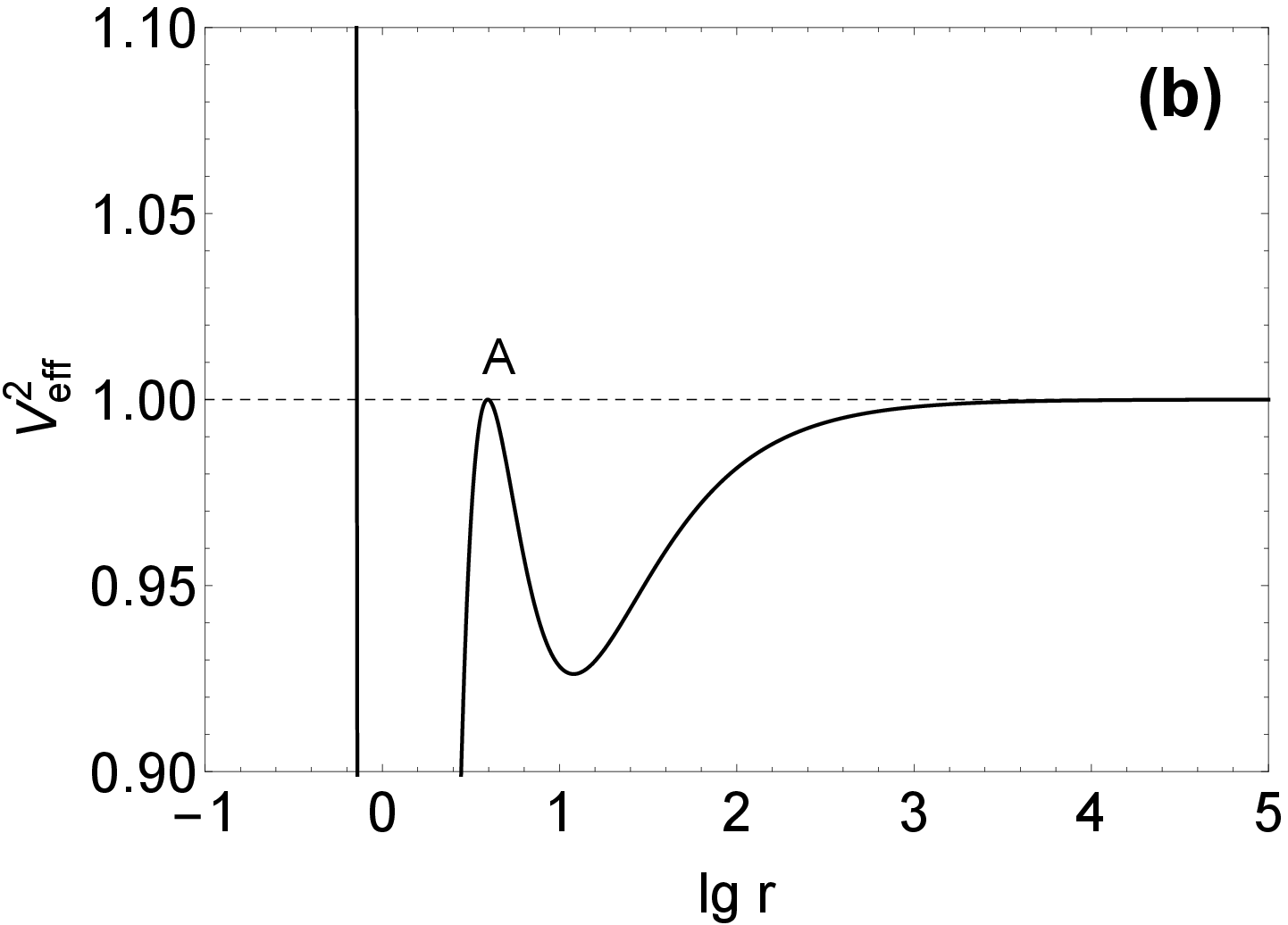}
     \vspace*{8pt}
\caption{The behavior of the effective potential for the massive particle with $\ell$ = $0.60M$, $b$ = $4.016M$, $\alpha$ = 0.50, $\beta$ = 1.00 and $M$ = 1. \label{v3} }
\end{figure}
\begin{figure}[ht!]
\centering
    \includegraphics[angle=0, width=0.4\textwidth]{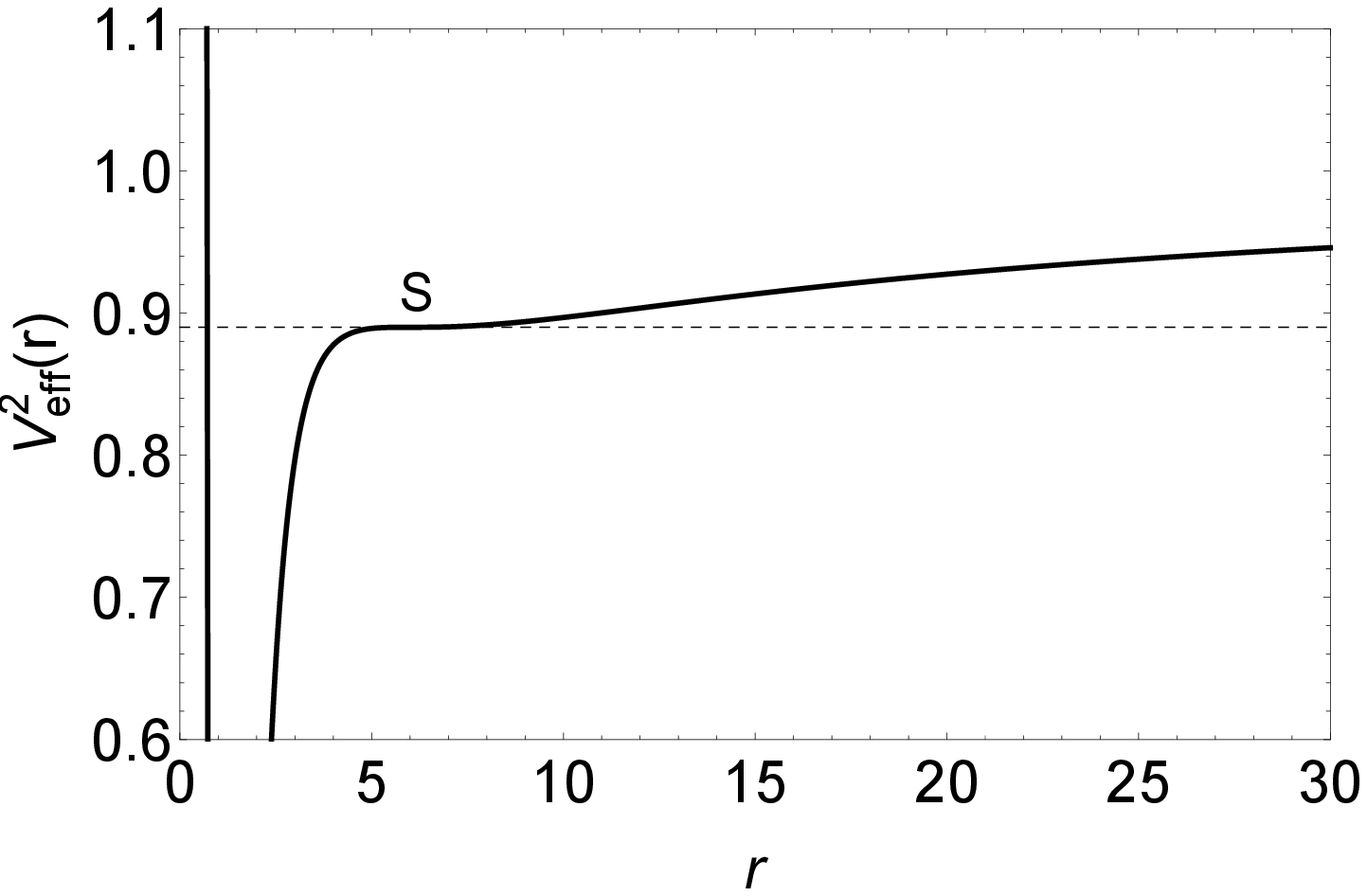}
     \vspace*{8pt}
\caption{The behavior of the effective potential for the massive particle with $\ell$ = $0.60M$, $b$ = 3.493$M$, $\alpha$ = 0.50, $\beta$ = 1.00, $M$ = 1 and $r_{S}$ = 5.997. \label{v4} }
\end{figure}

In Fig. \ref{v2}, we describe the influence of parameters ($\ell$, $b$, $\alpha$ and $\beta$) on the effective potential for time-like geodesic.
Form Fig. \ref{v2}(a), it can be found that for different $\ell$, the number of the effective potential curves and $r$-axis intersections contain three kinds.
We can see that the number of the effective potential curve and the $r$-axis common point reduces from 2 to 0 with the increase of $\ell$. There are three kinds of situations, which are no intersection, one intersection and double intersections.
Types of intersections in Fig. \ref{v2}(a) correspond to no horizon, one horizon and double horizons, respectively. This shows that the effective potential of the horizon is zero. Combining Figs. \ref{v2}(b), (c) and (d), we obtain that the stability of the orbital motion is associated with parameters $b$, $\alpha$ and $\beta$.
The peak of the effective potential increases with the increase of $b$ and decreases with the increases of $\alpha$ and $\beta$. The parameter $b$ have a conspicuous effect on the effective potential than parameters $\alpha$ and $\beta$. From Fig. \ref{v2}(b), we find that the values of $b$ can determine whether there exist escape orbits.
We also speculate that when the values of $b$ is in a certain range, the orbital motion contains only unstable orbital type, and being pulled into the black hole is the ultimate destination of the particle.
The limit values of the escape orbits, the extreme values of the stable orbits and the unstable orbits are respectively dealt in detail in Figs. \ref{v3} and \ref{v4}.

Through solving the equation $\frac{dV^{2}_{eff}}{dr}$ = 0 and adopting the control variable method, we obtain a limit value $b$ = $4.016M$ ($\beta$ = $1.00$, $\alpha$ = 0.50).
We plot the corresponding effective potential curves. They are illustrated in Fig. \ref{v3}. When $\alpha$ = 0.50, $\beta$ = 1.00 and $b$ $<$ $4.016M$, there are no escape orbits.

We derive the first derivative of the effective potential as
\begin{eqnarray}
&\frac{dV^{2}_{eff}}{dr}=&\frac{3\alpha^{2} \beta Mr^{2}(1+\frac{b^{2}}{r^{2}})(1-\frac{2Mr^{2}}{2M\ell^{2}+r^{3}})}{(\beta M+\alpha r^{3})^{2}}+(1+\frac{b^{2}}{r^{2}})(\frac{6Mr^{4}}{(2M\ell^{2}+r^{3})^{2}}-\frac{4Mr}{2M\ell^{2}+r^{3}})\nonumber \\
&&\times(1-\frac{\alpha \beta M}{\beta M+\alpha r^{3}})-\frac{2b^{2}(1-\frac{2Mr^{2}}{2M\ell^{2}+r^{3}})(1-\frac{\alpha \beta M}{\beta M+\alpha r^{3}})}{r^{3}},
\label{e12}
\end{eqnarray}
the second order derivative is
\begin{eqnarray}
&\frac{d^{2}V^{2}_{eff}}{dr^{2}}=&\frac{6\alpha^{2}\beta Mr^{2}(1+\frac{b^{2}}{r^{2}})(\frac{6Mr^{4}}{(2M\ell^{2}+r^{3})^{2}}-\frac{4Mr}{2M\ell^{2}+r^{3}})}{(\beta M+\alpha r^{3})^{2}}-\frac{18\alpha^{3}\beta M r^{4}(1+\frac{b^{2}}{r^{2}})(1-\frac{2M r^{2}}{2M\ell^{2}+r^{3}})}{(\beta M+\alpha r^{3})^{3}}\nonumber \\
&&-\frac{12\alpha^{2}\beta b^{2}M(1-\frac{2M r^{2}}{2M\ell^{2}+r^{3}})}{r(\beta M+\alpha r^{3})^{2}}+\frac{6\alpha^{2}\beta Mr(1+\frac{b^{2}}{r^{2}})(1-\frac{2Mr^{2}}{2M\ell^{2}+r^{3}})}{(\beta M+\alpha r^{3})^{2}}\nonumber \\
&&+(1+\frac{b^{2}}{r^{2}})(-\frac{36Mr^{6}}{(2M\ell^{2}+r^{3})^{3}}+\frac{36Mr^{3}}{(2M\ell^{2}+r^{3})^{2}}-\frac{4M}{2M\ell^{2}+r^{3}})(1-\frac{\alpha \beta M}{\beta M+\alpha r^{3}})\nonumber \\
&&-\frac{4b^{2}(\frac{6Mr^{4}}{(2M\ell^{2}+r^{3})^{2}}-\frac{4Mr}{2M\ell^{2}+r^{3}})(1-\frac{\alpha \beta M}{\beta M+\alpha r^{3}})}{r^{3}}+\frac{6b^{2}(1-\frac{2Mr^{2}}{2M\ell^{2}+r^{3}})(1-\frac{\alpha \beta M}{\beta M+\alpha r^{3}})}{r^{4}}
\label{e13}
\end{eqnarray}

Similarly, the effective potential of regular Hayward black hole is defined as
\begin{eqnarray}
&V^{2}_{eff}=\ (1-\frac{2M r^{2}}{r^{3}+2M \ell^{2}})(1+\frac{b^{2}}{r^{2}}).
\label{e14}
\end{eqnarray}
Differentiating Eq. (\ref{e14}) with respect to $r$, we have
\begin{eqnarray}
&\frac{dV^{2}_{eff}}{dr}=&\ (1+\frac{b^{2}}{r^{2}})(\frac{6Mr^{4}}{(2M\ell^{2}+r^{3})^{2}}-\frac{4Mr}{2M\ell^{2}+r^{3}})-\frac{2b^{2}}{r^{3}}(1-\frac{2Mr^{2}}{2M\ell^{2}+r^{3}}).
\label{e15}
\end{eqnarray}
Differentiating Eq. (\ref{e15}), we get the second order derivative of the effective potential
\begin{eqnarray}
&\frac{d^{2}V^{2}_{eff}}{dr^{2}}= &(1+\frac{b^{2}}{r^{2}})(-\frac{36Mr^{6}}{(2M\ell^{2}+r^{3})^{3}}+\frac{36Mr^{3}}{(2M\ell^{2}+r^{3})^{2}}-\frac{4M}{2M\ell^{2}+r^{3}})-\frac{4b^{2}}{r^{3}}(\frac{6Mr^{4}}{(2M\ell^{2}+r^{3})^{2}}\nonumber\\
&&-\frac{4Mr}{2M\ell^{2}+r^{3}})+\frac{6b^{2}}{r^{4}}(1-\frac{2Mr^{2}}{2M\ell^{2}+r^{3}}).
\label{e16}
\end{eqnarray}
Eqs. (\ref{e12}) and (\ref{e13}) when turning off $\alpha$ or $\beta$ will coincide with Eqs. (\ref{e15}) and (\ref{e16}), respectively.

From Eqs. (\ref{e15}) and (\ref{e16}), we obtain the radius of ISCO $r_{min}$ = 5.762 which corresponds to $b$ = 3.427$M$ and $\ell$ = $0.60M$ in regular Hayward black hole space-time.
Through analyzing Eqs. (\ref{e12}) and (\ref{e13}), if we take $\ell$= 0.6$M$, $\alpha$ = 0.5 and $\beta$ = 1.00, we could obtain an extreme value of angular momentum $b$ = $3.493M$ and the radius of ISCO $r_{S}$ = 5.997 in modified Hayward black hole space-time. It is shown in Fig. \ref{v4}.
This result shows that the introduction of the modified term makes the radius of ISCO and the corresponding angular momentum larger. Control of a single variable, when $\alpha$ = 0.50, $\beta$ = 1.00 and $b$ $<$ $3.493M$, the orbital types only have unstable orbits.


The corresponding equation of orbital motion can be written as
\begin{eqnarray}
&\dot{r}^{2}=\frac{E^{2}}{f(r)}-(1+\frac{b^{2}}{r^{2}})B(r).
\label{e17}
\end{eqnarray}
Combining the equation $\dot{\phi}$ = $\frac{b}{r^{2}}$ and replacing $R$ with $1/r$, Eq. (\ref{e17}) can be rewritten as
\begin{eqnarray}
 &(\frac{dR}{d\phi})^{2}=\frac{E^{2}}{f(R)b^{2}}-(\frac{1}{b^2}+R^2)B(R), \nonumber \\
&B(R)=1-\frac{2MR}{1+2Ml^{2}R^{3}}, \ \ \ f(R)=1-\frac{\alpha \beta M R^{3}}{\alpha + \beta M R^{3}}.
\label{e18}
\end{eqnarray}
Differentiating Eq. (\ref{e18}), the second order motion equation is written as
\begin{eqnarray}
&\frac{d^{2}R}{d\phi^{2}}=&\frac{E^{2}}{b^{2}} \frac{3 M \alpha^{2} \beta R^{2}}{2(\alpha+\beta M R^{3} - \alpha \beta M R^{3})^{2}}+ \frac{3M b^{2} R^{2}+3M}{b^{2}(1+2M \ell^{2}R^{3})^{2}}-\frac{2M}{b^{2}(1+2M\ell^{2}R^{3})} - R
\label{e19}
\end{eqnarray}

\subsection{Analyzing trajectory of a massive particle}
\begin{figure}[ht!]
\centering
    \includegraphics[angle=0, width=0.4\textwidth]{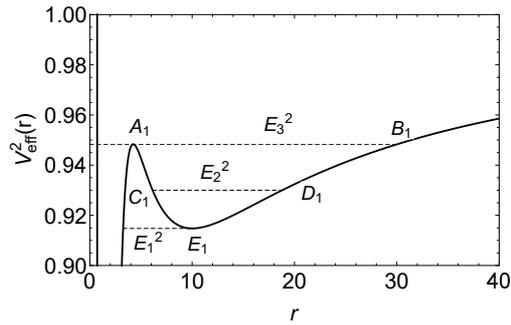}
     \vspace*{8pt}
\caption{The curve of the effective potential for the massive particle with $\ell$ = $0.60M$, $b$ = $3.80M$, $\alpha$ = 0.50, $\beta$ = 1.00 and $M$ = 1.  \label{v5} }
\end{figure}

\begin{figure}[ht!]
\centering
   \includegraphics[angle=0, width=0.3\textwidth]{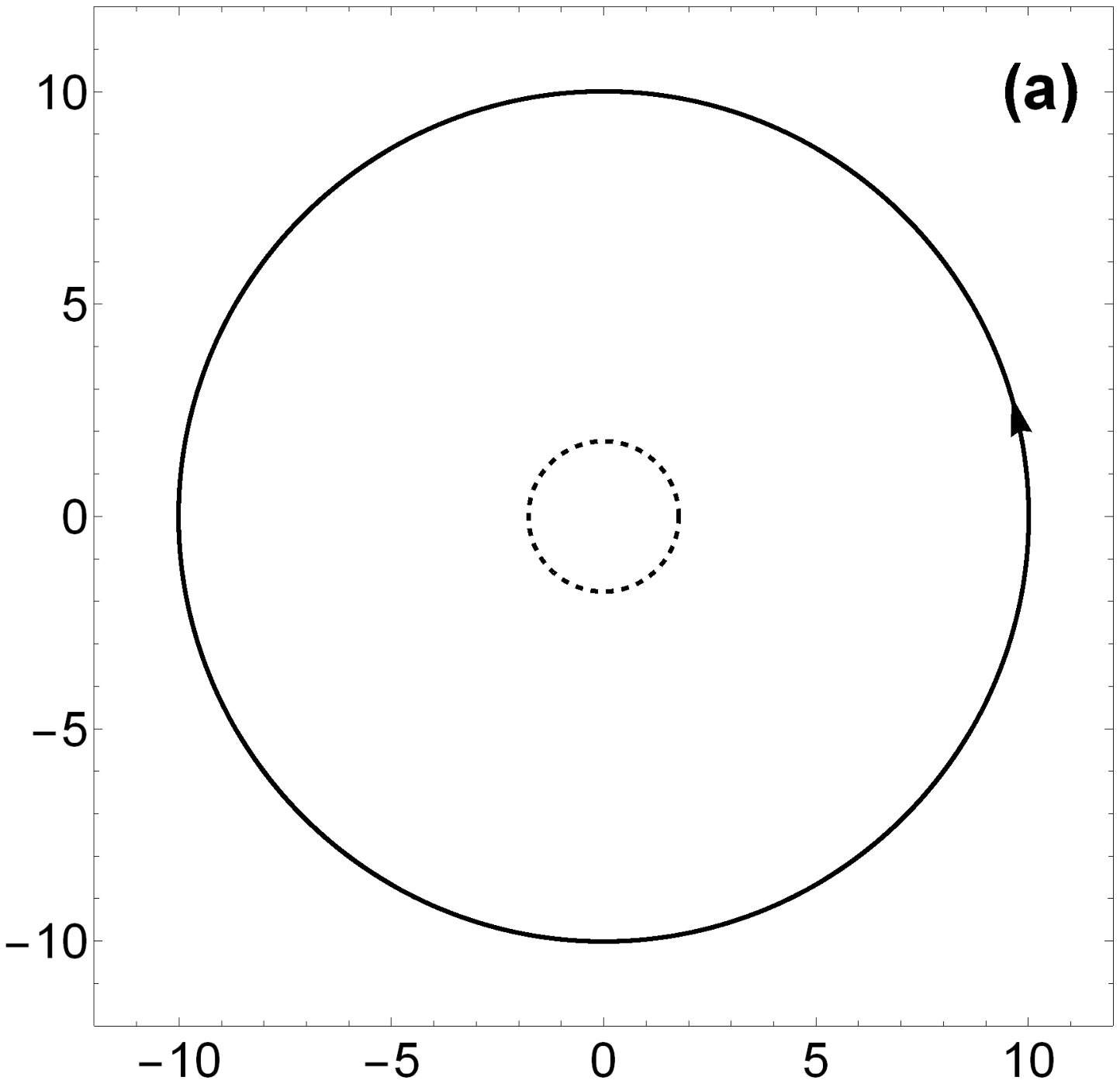}
  \ \ \includegraphics[angle=0, width=0.295\textwidth]{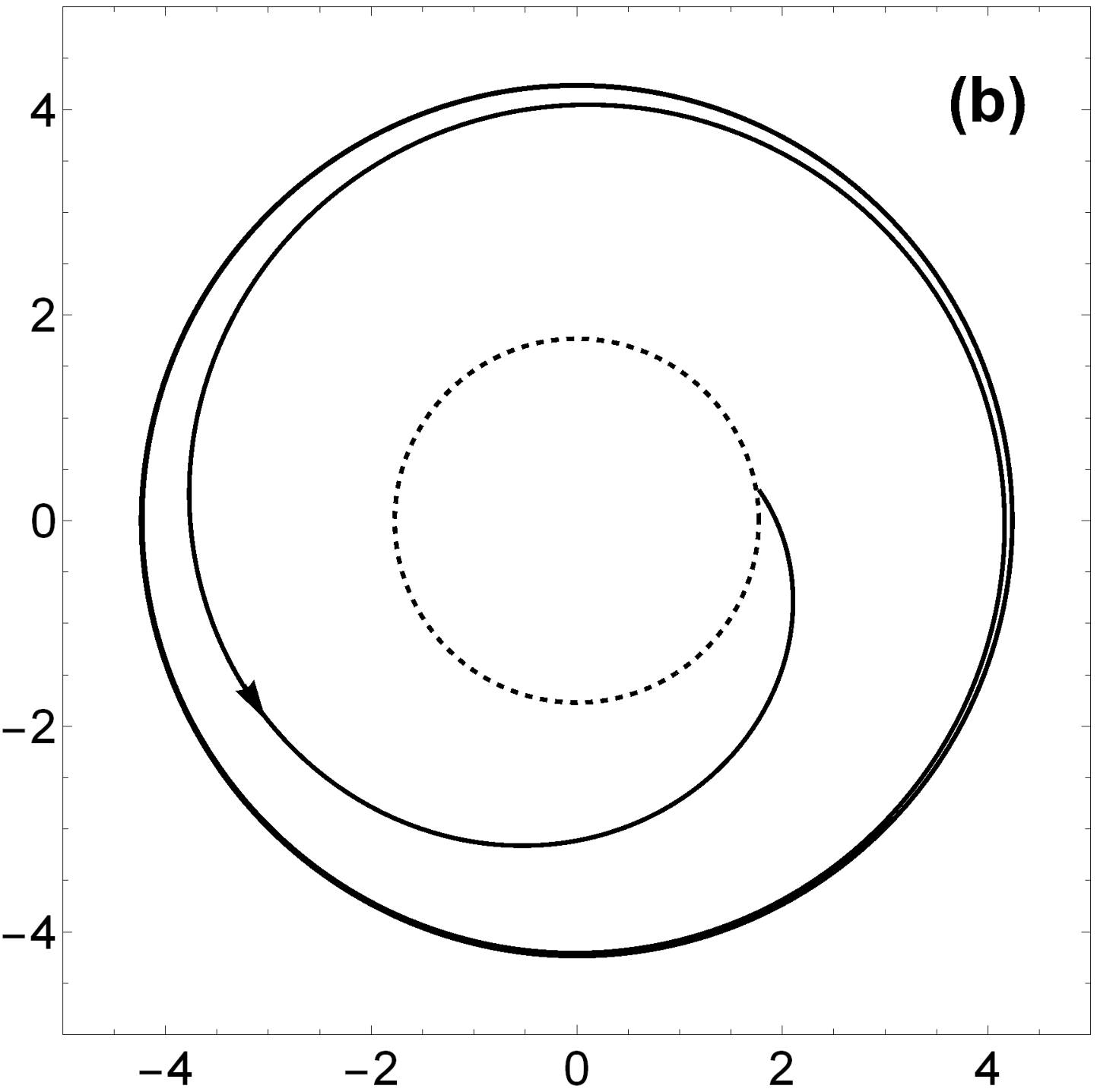} \\
   \includegraphics[angle=0, width=0.296\textwidth]{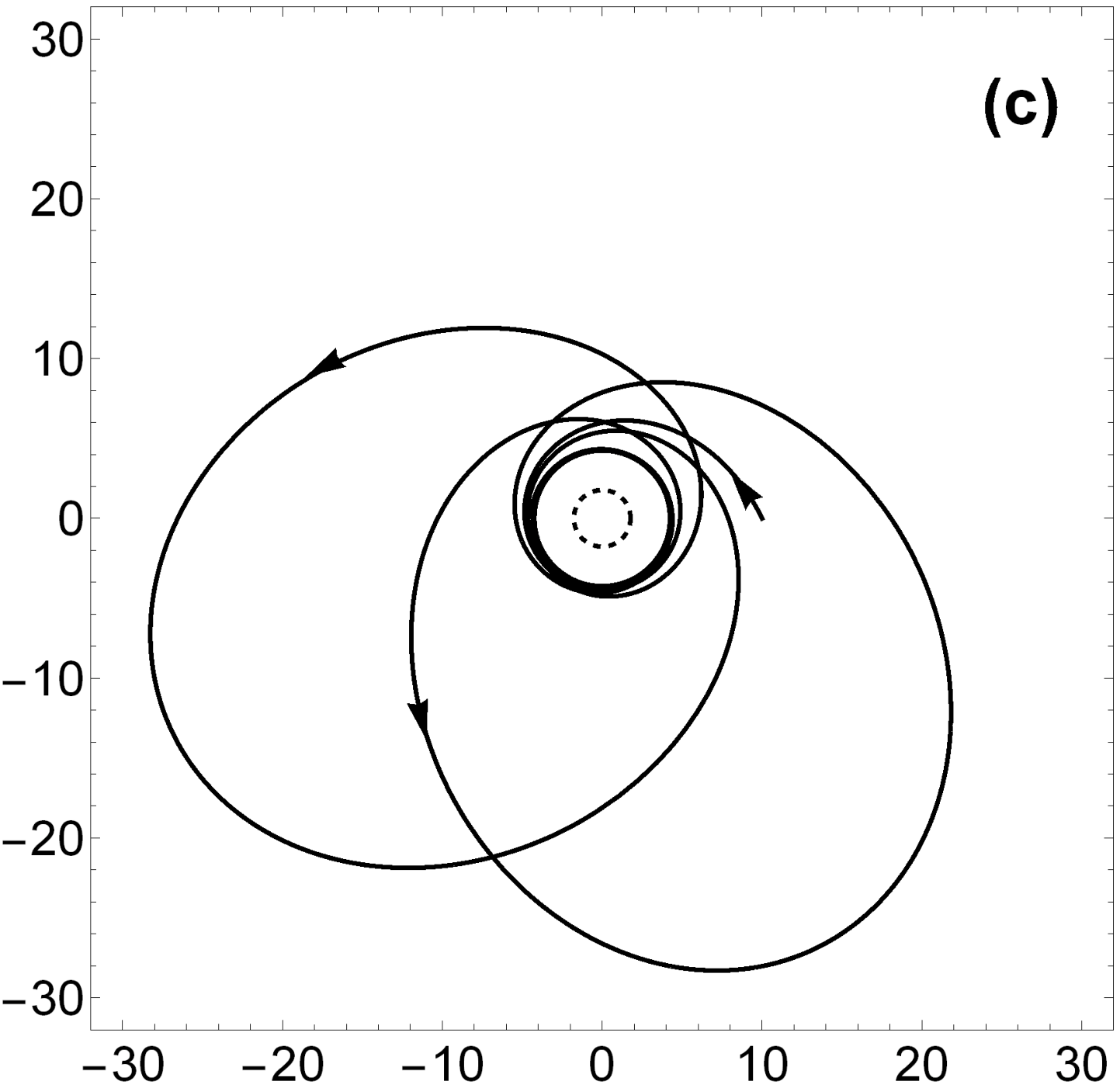}
  \ \includegraphics[angle=0, width=0.303\textwidth]{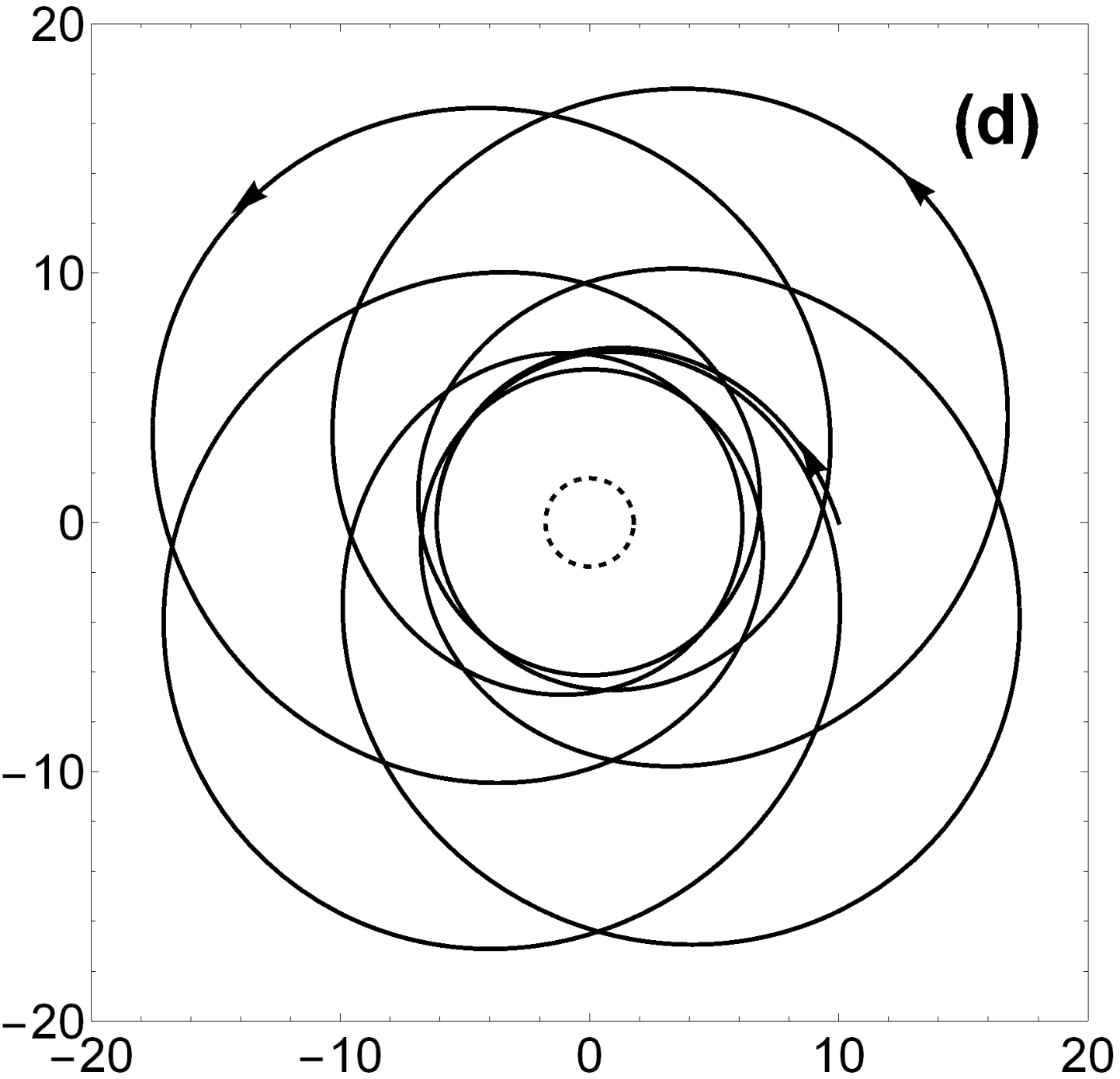}\\
   \includegraphics[angle=0, width=0.3\textwidth]{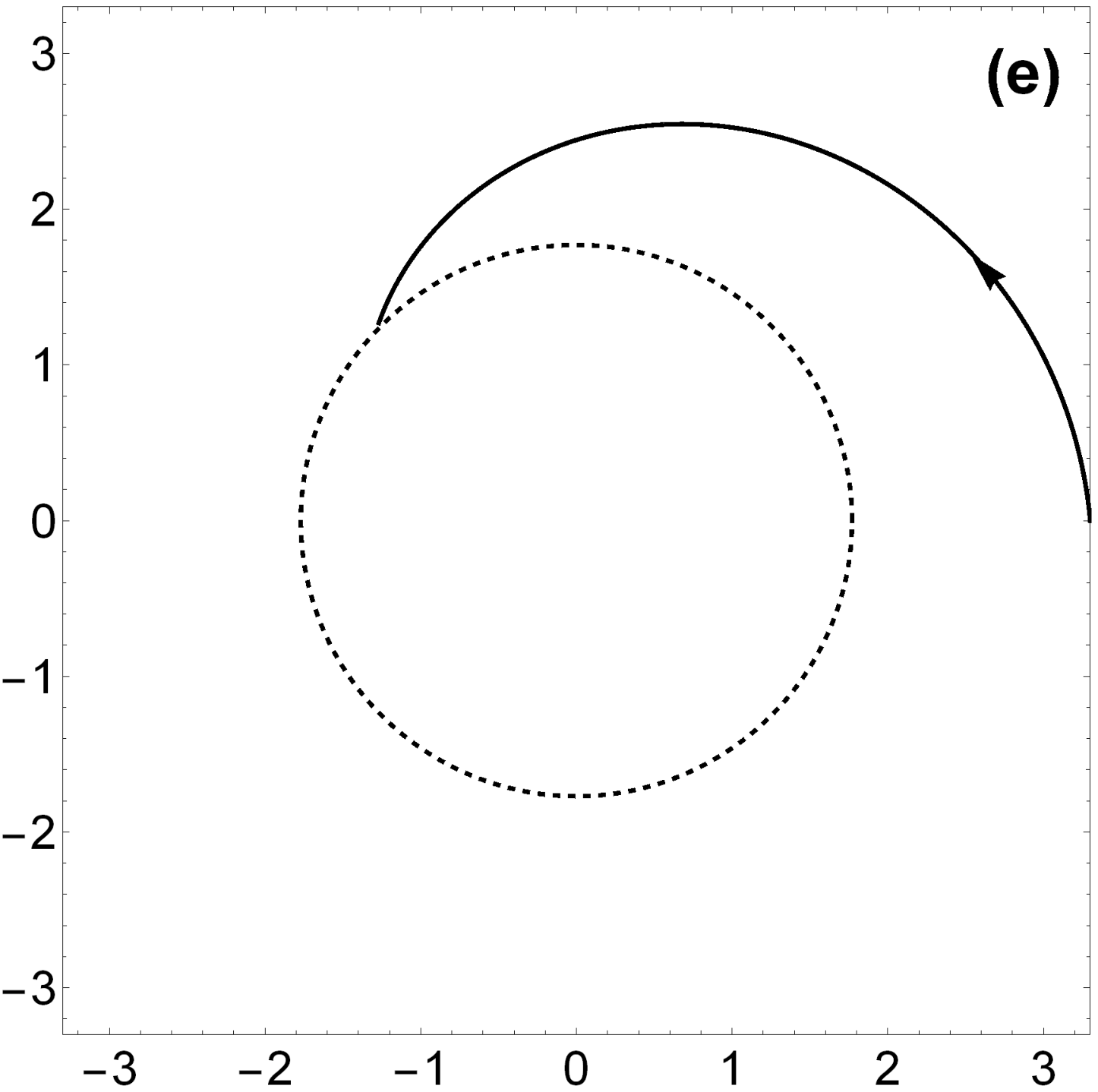}
   \vspace*{8pt}
\caption{The trajectories of the test particle with $\ell$ = $0.60M$, $b$ = $3.80M$, $\alpha$ = 0.50, $\beta$ = 1.00 and $M$ = 1 in modified Hayward black hole space-time.  \label{v6} }
\end{figure}

\begin{figure}[ht!]
\centering
  \includegraphics[angle=0, width=0.45\textwidth]{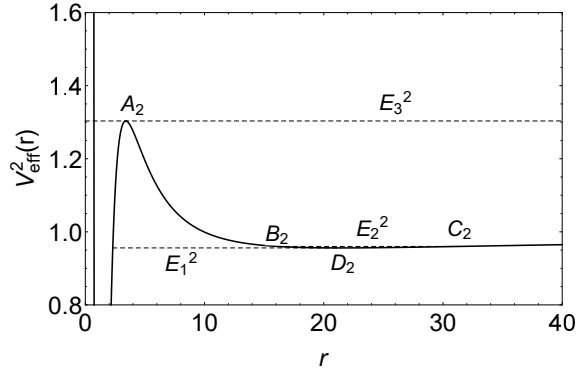}
  \vspace*{8pt}
\caption{The curve of the effective potential for the massive particle with $\ell$ = $0.60M$, $b$ = $5.00M$, $\alpha$ = 0.50, $\beta$ = 1.00 and $M$ = 1. \label{v7} }
\end{figure}

\begin{figure}[ht!]
\centering
  \includegraphics[angle=0, width=0.31\textwidth]{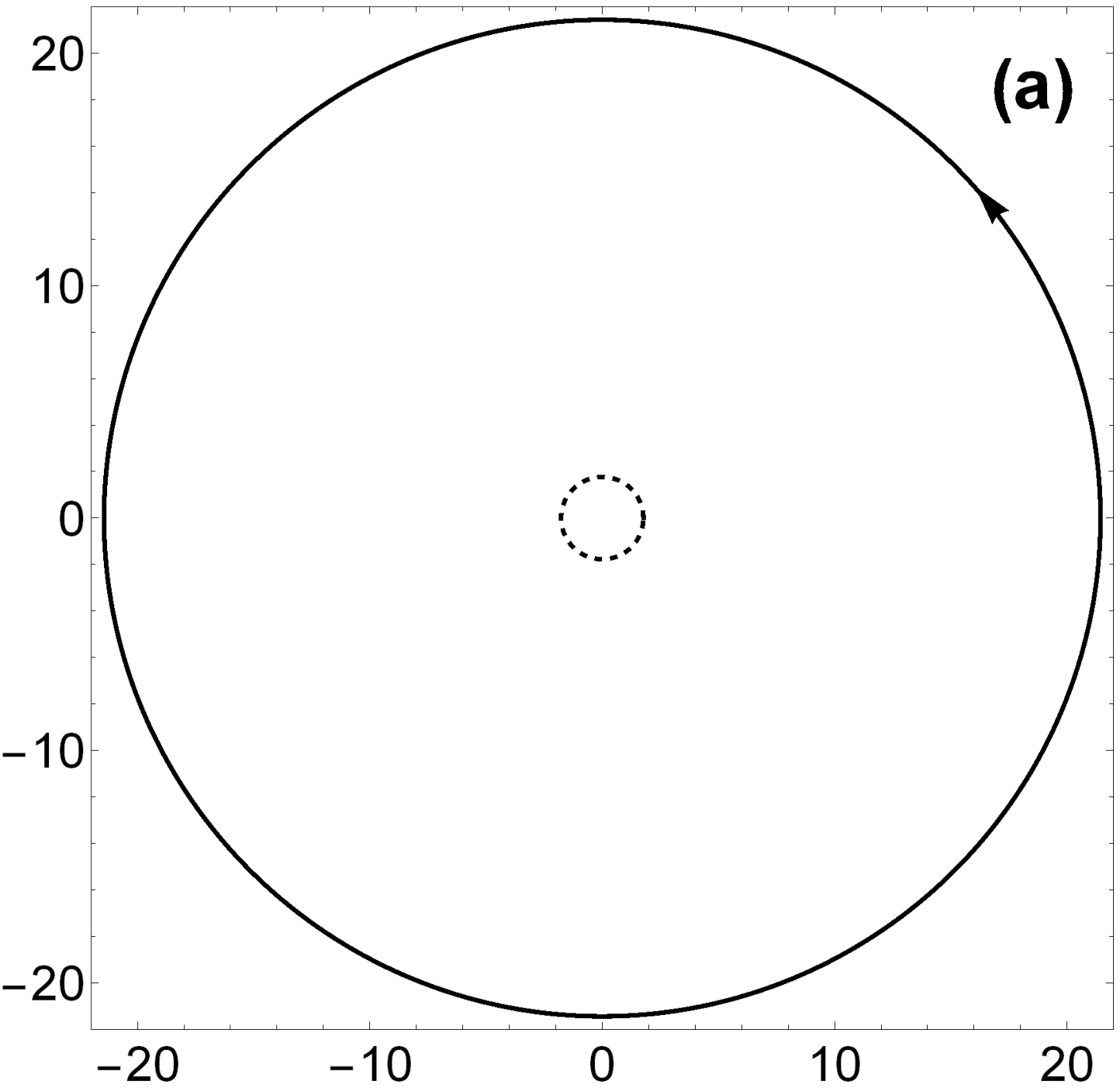} \
  \includegraphics[angle=0, width=0.305\textwidth]{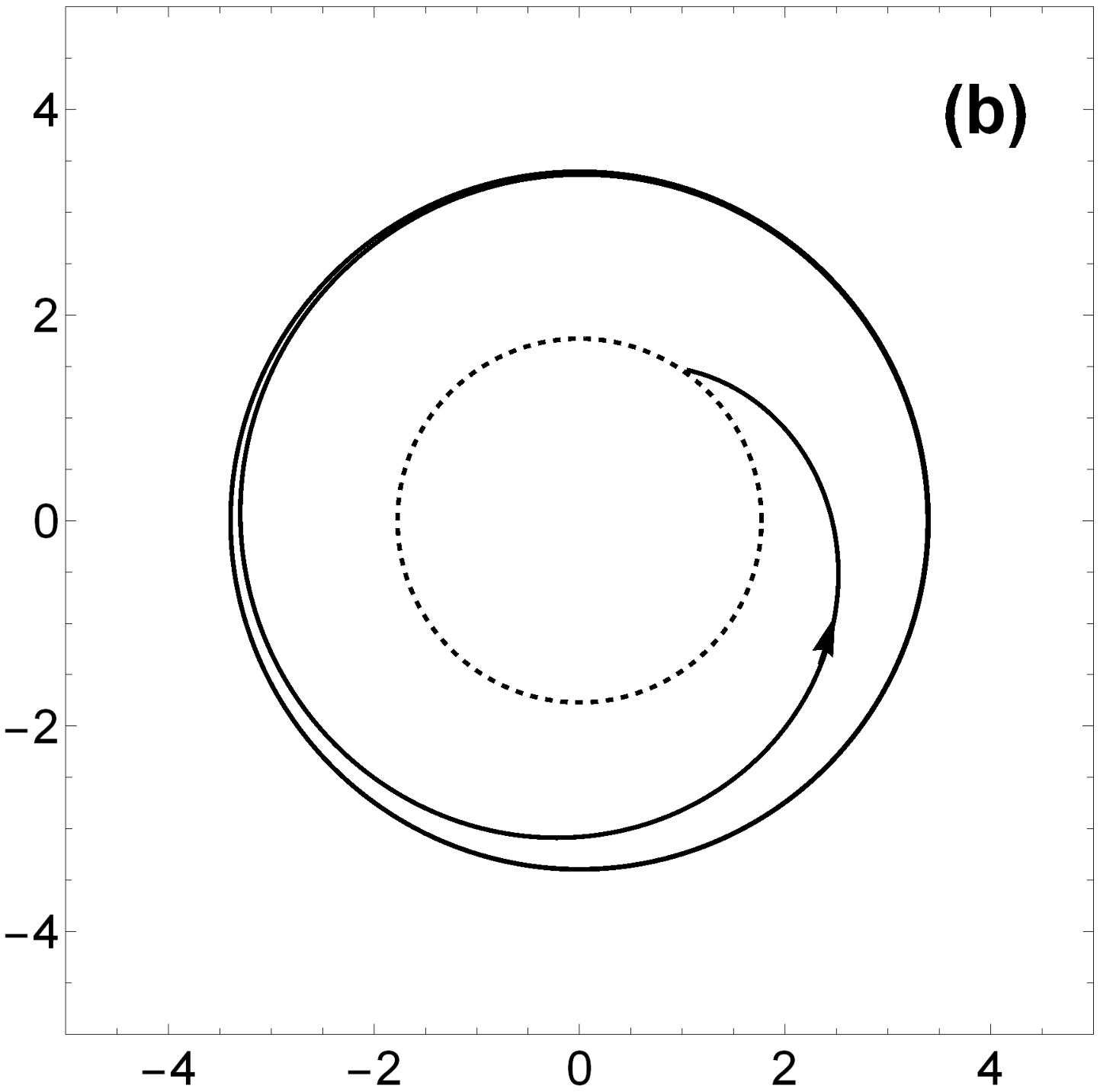} \\
  \includegraphics[angle=0, width=0.313\textwidth]{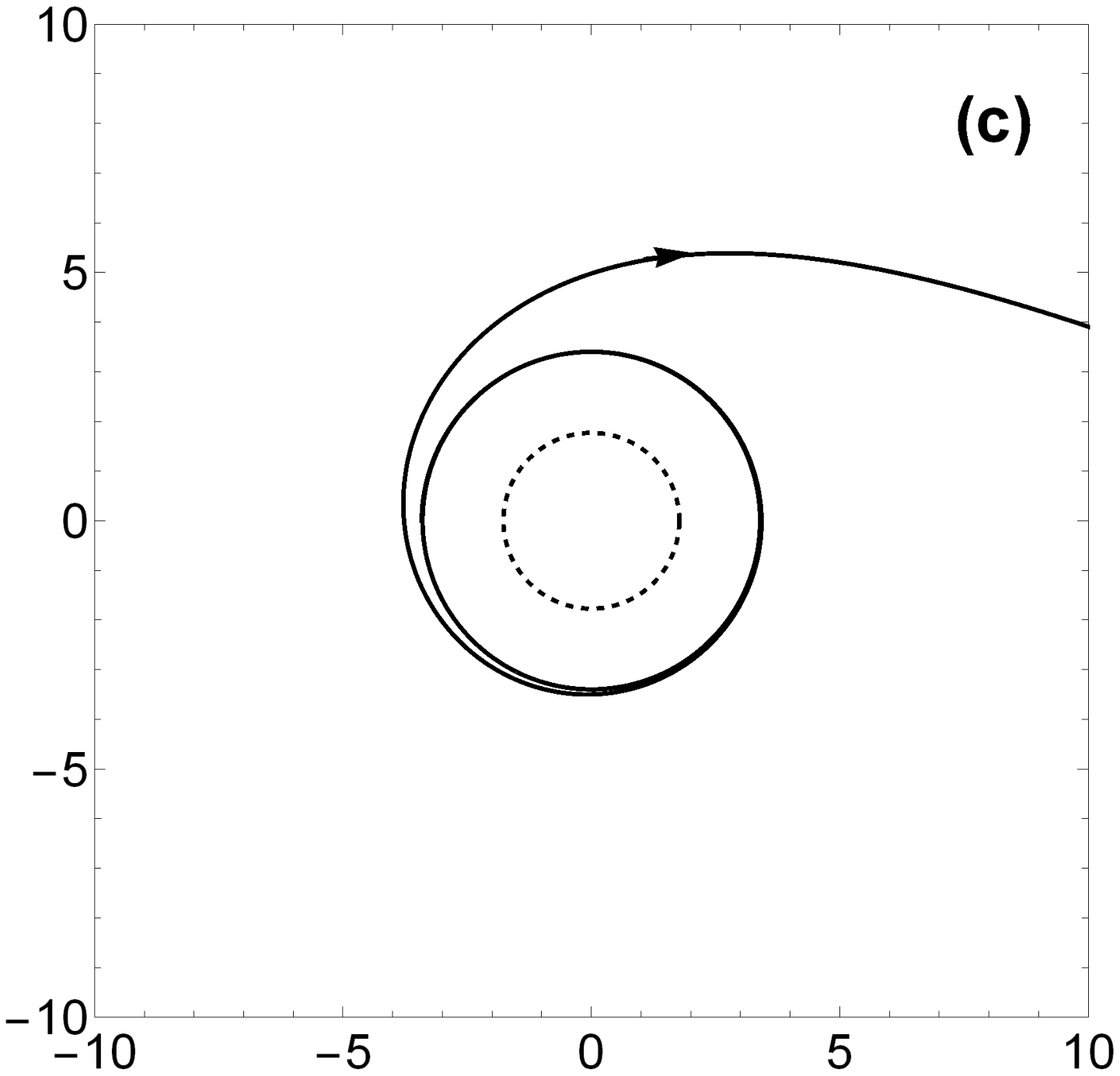}
  \includegraphics[angle=0, width=0.307\textwidth]{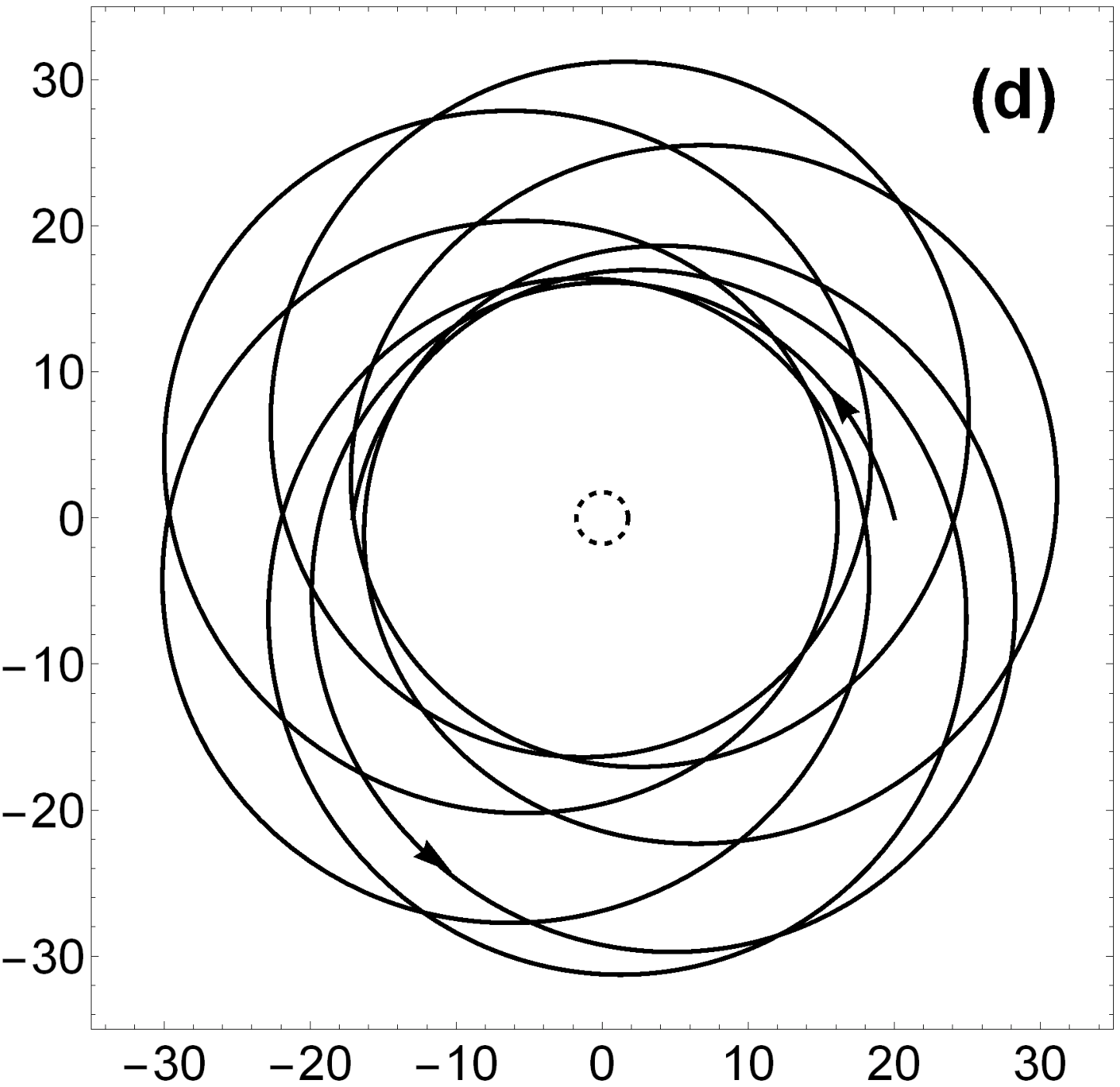} \\ \ \
  \includegraphics[angle=0, width=0.302\textwidth]{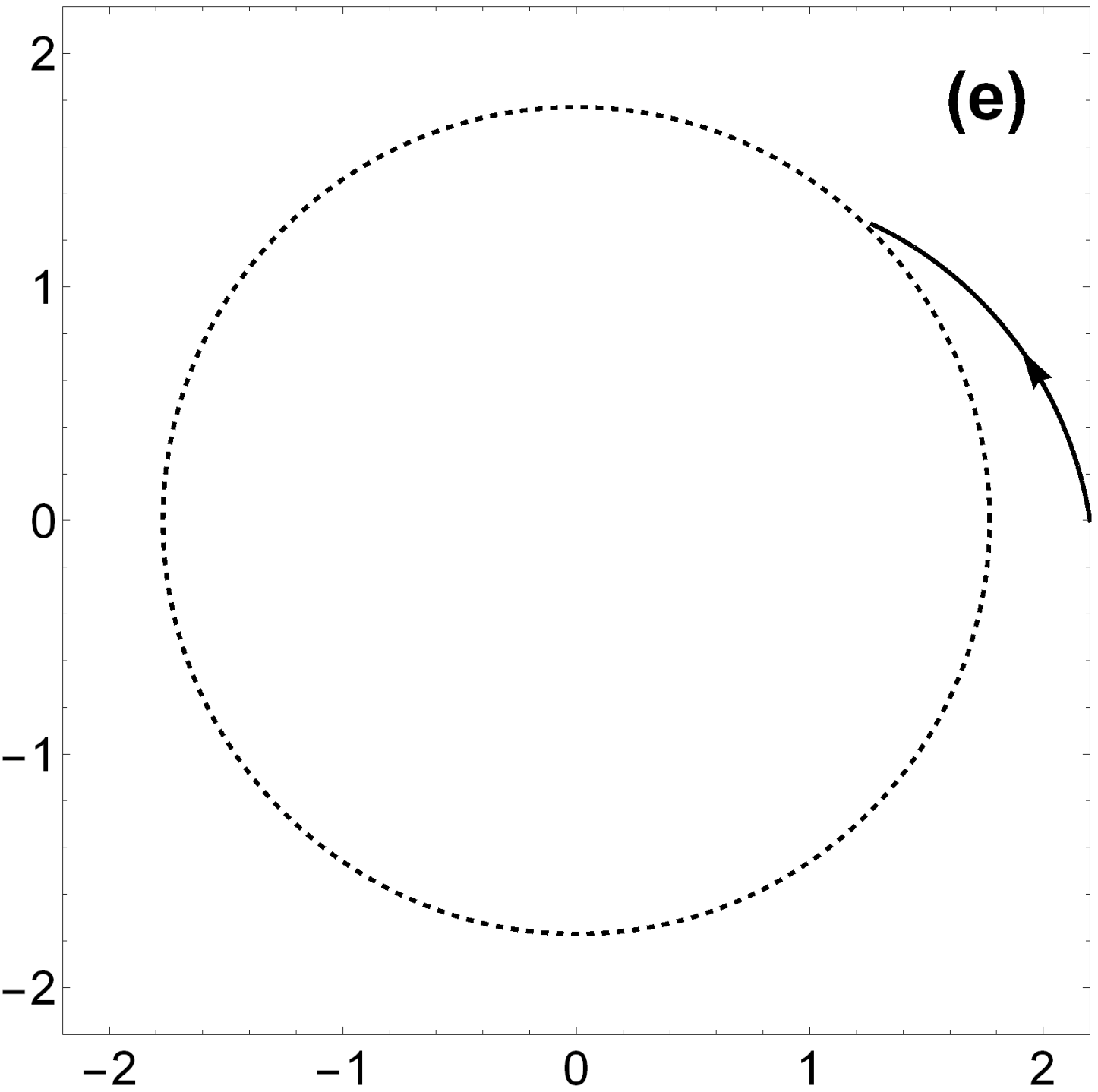}
  \includegraphics[angle=0, width=0.312\textwidth]{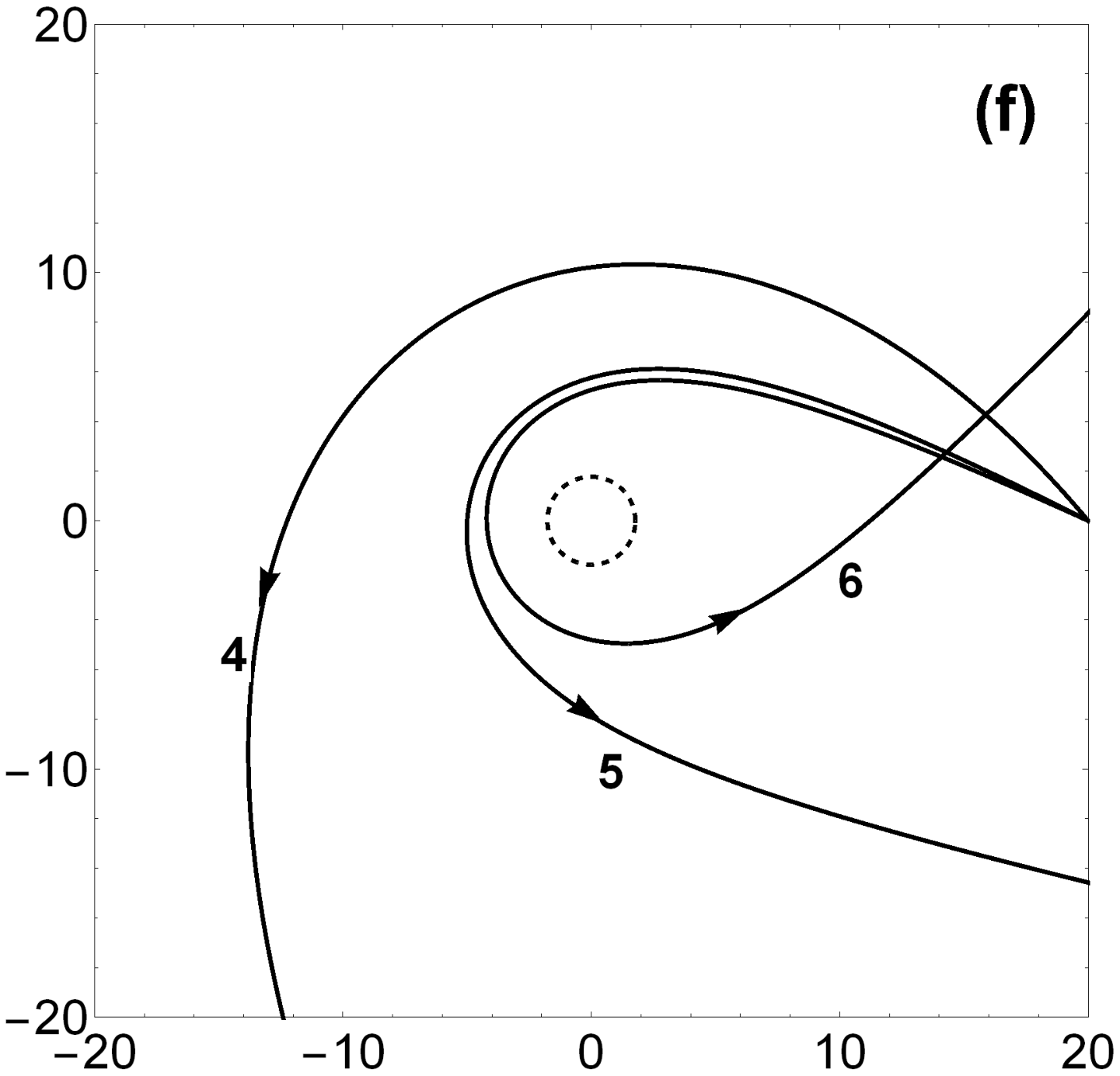}
     \vspace*{8pt}
\caption{The trajectories of the test particle with $\ell$ = $0.60M$, $b$ = $5.00M$, $\alpha$ = 0.50, $\beta$ = 1.00 and $M$ = 1 in modified Hayward black hole space-time. \label{v8} }
\end{figure}

We discover all possible orbits of a massive particle in modified Hayward black hole space-time by solving Eqs. (\ref{e18}) and (\ref{e19}) numerically. In this paper, we only discuss the orbital motion of a massive particle thrown in the direction of the black hole.
In Figs. \ref{v5} to \ref{v8}, we plot the effective potential and all possible orbits with different $b$, and once again the effect of $b$ on orbital stability was verified.

In Fig.\ref{v6}, when $E^{2}$ = $E^{2}_{1}$ = 0.915, the test particle moves on a stable circular orbit with a radius of $r_{E_{1}}$ = 10.01, it is shown in Fig. \ref{v6}(a).
When $E^{2}$ = $E^{2}_{2}$ = 0.93, the test particle moves on bound orbits with radius between $r_{C_{1}}$ = 6.93 and $r_{D_{1}}$ = 18.80.
The precession direction of the bound orbit is clockwise and the precession speed is fast. $C_{1}$ is the perihelion while $D_{1}$ is the aphelion, as shown in Fig. \ref{v6}(d).
When $E^{2}=E^{2}_{3}=0.948$, the test particle can move on an unstable circular orbit, with radius of $r_{A_{1}}$ = 4.24. Any small disturbance will make the test particle out of their original orbit.
Then there are two choices. One is falling into the black hole, another is moving on an unstable precession orbit between the radius of $r_{A_{1}}$ and $r_{B_{1}}$ = 30.04, these cases are described in Figs. \ref{v6}(b) and (c), respectively.
The choice of the test particle depends on how the energy level of the test particle changes. In addition, when 0 $<$ $E^{2}$ $<$ $E^{2}_{3}$, if we throw the test particle from anywhere in the left area of peak $A_{1}$, it will eventually fall into the black hole, as shown in Fig. \ref{v6}(e).

In Fig. \ref{v7}, when $E^{2}$ = $E^{2}_{1}$ = 0.9559, the test particle will move on a stable circular orbit with radius $r_{D_{2}}$ = 21.44, it is shown in Fig. \ref{v8}(a).
Being compared to Fig. \ref{v6}(a), we find that with the increase of $b$, the radius of the stable circular orbit increases, i.e., $r_{E_{1}}$ $<$ $r_{D_{2}}$.
If we choose $E^{2}$ = $E^{2}_{2}$ = 0.96, the test particle will move on bound orbit between the radius of $r_{B_{2}}$ = 16.12 and $r_{C_{2}}$ = 31.37. $r_{B_{2}}$ is the perihelion radius while $r_{C_{2}}$ is the aphelion radius, as shown in Fig. \ref{v8}(d).
The precession direction of the bound orbit is counterclockwise which is different from Fig. \ref{v6}(d), and the precession velocity of Fig. \ref{v8}(d) is slower than Fig. \ref{v6}(d).
When $E^{2}$ = $E^{2}_{3}$ = 1.30, the test particle will move on an unstable circular orbit with the radius of $r_{A_{2}}$ = 3.40, it is shown in Figs. \ref{v8}(b) and (c).
The radius of unstable circular orbit decreases with the increase of $b$. As $E^{2}_{1}$ $<$ $E^{2}$ $<$ $E^{2}_{3}$, the test particle may also be absorbed by the black hole, but its range becomes smaller due to the left movement of the peak of the effective potential, as shown in Fig. \ref{v8}(e).
For $E^{2}_{1}$ $<$ $E^{2}$ $<$ $E^{2}_{3}$, it can also move on escape orbits, then will get rid of the black hole and escape to infinity. We choose three energy levels (1 $<$ $E^{2}_{4}$ $<$ $E^{2}_{5}$ $<$ $E^{2}_{6}$) between $E^{2}_{1}$ and $E^{2}_{3}$ to plot the trajectories of the test particle in Fig. \ref{v8}(f).
The results show that the orbital curvature with a higher energy level is higher, i.e., $C_{4}$ $<$ $C_{5}$ $<$ $C_{6}$.
The comparison between Figs. \ref{v6} and \ref{v8} also validates our previous view when $b$ is too small, there is no escape orbit.
We also find that for the stable and unstable circular orbits, the energy levels of the orbits increase with the increases of $b$.

\

Combining Figs. \ref{v5} to \ref{v8}, we find that the parameters $b$ do have an observable effect on the properties of orbital motion in modified Hayward black hole space-time.
  The radius of the stable and unstable circular orbits are related to parameter $b$. With the increase of $b$, the radius of stable circular orbits increase while the radius of unstable circular orbits decrease, their energy levels both increase. For bound orbits, the precession direction and velocity of bound orbits are also associated with parameters $b$. There is no escape orbits for $\ell$=0.60$M$, $b$=3.80$M$, $\alpha$=0.50 and $\beta$=1.00$M$. $b$=3.80$M$ $<$ $4.02M$ is too small to produce escape orbits.

\section{Conclusion}
In this paper, we have investigated the trajectories of massive particle by resolving the equation of orbital motion numerically. There exist four orbital types: stable and unstable circular orbits, planetary orbits, absorbing orbits and escape orbits in time-like geodesic of modified Hayward black hole space-time. By analyzing the behavior of effective potential and plotting all possible orbits, the properties of orbital motion are discussed in detail. It is obtained that parameters $b$ have a conspicuous effect on the effective potential than parameters $\alpha$ and $\beta$. The influence of parameters $b$, $\alpha$ and $\beta$ on the peak of the effective potential is different. The peak of the effective potential increases with $b$ increasing and decreases with $\alpha$ and $\beta$ increasing. We have found the limit condition of escape orbit and ISCO for fixed values of parameters. Through plotting the trajectories of a massive particle, we have discovered that the precession direction and velocity of bound orbits are related to the parameters $b$. If $\beta$ = 1.00, when $b$ changes from $3.80M$ to $5.00M$, the precession direction of the bound orbits changes from clockwise to counterclockwise and the velocity of the precession becomes slower. We also find that the influence of energy level on the same type of orbital motions is obvious. For the escape orbits, the higher the energy level, the larger the curvature of the orbit.

{\bf Acknowledgments}

This work was supported in part by the National Natural Science Foundation of China (Grant No. 11565016),
the Special Training Program for Distinguished Young Teachers of the Higher Education Institutions of Yunnan Province (Grant No. 1096837802),
the Applied Basic Research Programs of Yunnan Provincial Science and Technology Department (Grant No. 2016FB008).

\end{document}